# Attack on Fully Homomorphic Encryption over the Integers

Gu Chunsheng
School of Computer Engineering
Jiangsu Teachers University of Technology
Changzhou, China, 213001
guchunsheng@gmail.com

**Abstract:** This paper presents a heuristic attack on the fully homomorphic encryption over the integers by using lattice reduction algorithm. Our result shows that the FHE in [DGHV10] is not secure for some parameter settings. We also present an improvement scheme to avoid the lattice attack in this paper.

## 1. Introduction

Rivest, Adleman and Dertouzos [RAD78] introduced a notion of privacy homomorphism. But until 2009, Gentry [Gen09] constructed the first fully homomorphic encryptions based on ideal lattice, all previous schemes are insecure. Following the breakthrough of [Gen09], there is currently great interest on fully-homomorphic encryption [SV10, vDGHV10, SS10, GH11a, GH11b, BV11a, BV11b, BGV11, CJMNT11, CMNT11]. In these schemes, the simplest one is certainly the one of van Dijk, Gentry, Halevi and Vaikuntanathan [DGHV10]. The public key of this scheme is a list of approximate multiples $\{x_i = q_i p + 2r_i\}_{i=1}^{\tau}$ for an odd integer $p$, where $q_i, r_i$ is the uniform random integers over $Z$ such that $|r_i| < 2^{\lambda-1}$. The secret key is $p$. To encrypt a message bit $m$, the ciphertext is evaluated as $c = \sum_{i \in T, T \subseteq [\tau]} x_i + 2r + m$, where $|r| < 2^{\lambda-1}$. To decrypt a ciphertext, compute the message bit $m = [c]_p \bmod 2$, where $[c]_p$ is an integer in $(-p/2, p/2)$.

To conveniently compare, we simply describe the known attacks considering in the Section 5 and appendix B in [DGHV10]. Section 5 in [DGHV10] considered known attacks on the AGCD problem for two numbers $(x_0, x_1)$ and many numbers $(x_0, \cdots, x_t)$. These attacks mainly discussed how to solve approximate GCD problem, i.e. the secret key $p$.

The appendix B.1 in [DGHV10] analyzed Nguyen and Stern's orthogonal lattice attack. Given

$\vec{x}=(x_0,...,x_t)=p\vec{q}+\vec{r}$ , where $\vec{q}=(q_0,...,q_t)$ and $\vec{r}=(r_0,...,r_t)$ , consider the $t$-dimensional lattice $L_{\vec{x}}^{\perp}$ of integer vectors orthogonal to $\vec{x}$. It is easy to verify that any vector that is orthogonal to both $\vec{q}$ and $\vec{r}$, that is, is in the lattice $L_{\vec{q},\vec{r}}^{\perp}$, it is also in $L_{\vec{x}}^{\perp}$. According to [DGHV10], the idea of the attack is to reduce $L_{\vec{x}}^{\perp}$ to recover $t-1$ linearly independent vectors of $L_{\vec{q},\vec{r}}^{\perp}$, and further recover $\vec{q}$ and $\vec{r}$, and $p$. Then Dijk et al. discussed that when $t>\gamma/(\eta-\rho)$, lattice reduction algorithms can not find a $2^{\eta-\rho}$ approximate short vectors in $L_{\vec{q},\vec{r}}^{\perp}$ on the worst-case.

Dijk et al. also analyzed a similar above attack by using the constraint $x_i-r_i=0 \bmod p$, which also paid close attention to how to solve for the $\vec{r}$. They considered a lattice as follows.

$$M=\begin{pmatrix} x_1 & R_1 & & & \\ x_2 & & R_2 & & \\ & & & \ddots & \\ x_t & & & & R_t \end{pmatrix}.$$

But one needs to find $t$ linearly independent short vectors of the lattice $M$ to obtain the success of this attack. That is, each $l_1$ norm among $t$ vectors is at most $p/2$. When $t$ is large, solving these vectors is very difficult by using lattice reduction algorithm.

In addition, instead of applying linear system $x_i-r_i=0 \bmod p$, Coppersmith's method looks at quadratic system $(x_i-r_i)^2=0 \bmod p^2$ and $(x_i-r_i)(x_j-r_j)=0 \bmod p^2$, etc, and finds one of the $r_i$ and thereof $p$ and all other $r_i$'s by solving some small vectors in new lattice.

In a word, the attacks considering in the Section 5 and appendix B in [DGHV10] is how to recover the secret key $p$, and their security analysis depends on the worst-case performance of the currently known lattice reduction algorithms.

The lattice we constructed is very similar to the lattice $M$. However, our attack only requires find one short vectors with certain condition, and not to solve $t$ short vectors. Moreover, our attack merely recovers the plaintext from a ciphertext and depends upon the average-case performance of the lattice reduction algorithms. On the other hand, if suppose $\vec{x}=(c,x_0,...,x_t)=p\vec{q}+2\vec{r}+m$ with a ciphertext $c$, then our attack in some sense is similar to solving a short vector of orthogonal lattice $L_{\vec{q}}^{\perp}$, which is different from the lattices $L_{\vec{x}}^{\perp}$ or $L_{\vec{q},\vec{r}}^{\perp}$ considering in the Section 5 and appendix B in [DGHV10].

**Our Contribution.** Our main observation is that one can directly obtain the plaintext from a ciphertext and the public key without using the secret key for some parameter settings of the FHE in [DGHV10]. The attack in this paper is different from the known attacks considering in [DGHV10]. Because our method is how to recover the plaintext from a ciphertext, whereas the attacks they considered is how to solve the secret key in the scheme. So, our result shows that the FHE in [DGHV10] is not secure for some practical parameters.

**Organization of This Paper.** Section 2 gives some notations and definitions, and the lattice reduction algorithms. Section 3 constructs a new lattice based on the public key, and presents a polynomial time algorithm to directly obtain plaintext from ciphertext. Section 4 presents an improvement scheme. Section 5 concludes this paper.

## 2. Preliminaries

### 2.1 Notations

In this paper, we follow the parameter setting of [DGHV10]. Let $\lambda$ be a security parameter, $[\lambda]=\{1,...,\lambda\}$ be a set of integers. Let $\gamma$ be bit-length of the integers in the public key, $\eta$ the bit-length of the secret key, $\rho$ the bit-length of the noise, and $\tau$ the number of integers in the public key. To conveniently describe, we concretely set $\rho=\lambda$, $\eta=4\lambda^2$, $\gamma=\lambda^5$, and $\tau=\gamma+\lambda$ throughout this paper.

Let $w \xleftarrow{\Psi} S$ denote to choose an element $w$ in $S$ according to the distribution $\Psi$.

### 2.2 Lattices

A lattice in $\mathbb{R}^m$ is the set of all integral combination of $n$ linearly independent vectors $b_1,...,b_n$ in $\mathbb{R}^m$ ($m \geq n$), namely $L=L(b_1,...,b_n)=\{\sum_{i=1}^{n} x_i b_i, x_i \in Z\}$, usual denoted as a matrix $B$. Any such $n$-tuple of vectors $b_1,...,b_n$ is called a basis of the lattice $L$. Every lattice has an infinite number of lattice bases. Two lattice bases $B_1, B_2 \in \mathbb{R}^{n\times m}$ are equivalent if and only if $B_1 = UB_2$ for some unimodular matrix $U \in \mathbb{Z}^{n\times n}$. The volume of a lattice $L$ is the determinant of any basis of $L$, namely $vol(L)=\det(L)=\sqrt{B^T B}$.

### 2.3 Lattice Reduction Algorithm

Given a basis of the lattice $b_1,...,b_n$, one of the most famous problems of the algorithm theory of lattices is to find a short nonzero vector. Currently, there is no polynomial time algorithm for solving a shortest nonzero vector in a given lattice. The most celebrated LLL reduction finds a vector whose approximating factor is at most $2^{(n-1)/2}$. In 1987, Schnorr [Sch87] introduced a hierarchy of reduction concepts that stretch from LLL reduction to Korkine-Zolotareff reduction which obtains a polynomial time algorithm with $(4k^2)^{n/2k}$ approximating factor for lattices of any rank. The running time of Schnorr's algorithm is poly(size of basis)*HKZ(2*k*), where HKZ(2*k*) is the time complexity of computing a 2*k*-dimensional HKZ reduction, and equal to $O(k^{k/2+o(k)})$. If we use the probabilistic AKS algorithm [AKS01], HKZ(2*k*) is about $O(2^{2k})$.

**Theorem 2.1 (Sch87 Theorem 2.6)** Every block $2k$-reduced basis $b_1,...,b_{mk}$ of lattice $L$ satisfies $\|b_1\| \leq \sqrt{\gamma_k}\beta_k^{\frac{m-1}{2}}\lambda_1(L)$, where $\beta_k$ is another lattice constant using in Schnorr's analysis of his algorithm.

Shnorr [Sch87] showed that $\beta_k \leq 4k^2$, and Ajtai improved this bound to $\beta_k \leq k^{\varepsilon}$ for some positive number $\varepsilon > 0$. Recently, Gama Howgrave, Koy and Nguyen [GHKN06] improved the approximation factor of the Schnorr's 2*k*-reduction to $\|b_1\|/\lambda_1(L) \leq \sqrt{\gamma_k}(4/3)^{(3k-1)/4}\beta_k^{n/2k-1}$, and proved the following result via Rankin's constant.

**Theorem 2.2 (GHKN06 Theorem 2, 3)** For all $k \geq 2$, Schnorr's constant $\beta_k$ satisfies: $k/12 \leq \beta_k \leq (1+k/2)^{2\ln 2+1/k}$. Asymptotically it satisfies $\beta_k \leq 0.1 \times k^{2\ln 2+1/k}$. In particular, $\beta_k \leq k^{1.1}$ for all $k \leq 100$.

**Observation 2.3 (NS06).** For lattice $L$, the first vector $b_1$ output by LLL is satisfied to the ratio $\|b_1\|/\lambda(L) \approx (1.02)^n$ on the average.

# 3. Attack on FHE Scheme

To describe simplicity, we first refer the FHE scheme in [DGHV10], then construct a new lattice based on the public key and recover the plaintext bit from a ciphertext by applying LLL lattice reduction algorithm.

### 3.1 Fully Homomorphic Encryption

**KeyGen($\lambda$).** The secret key is a random odd $\eta$-bit integer: $p \xleftarrow{\Psi} (2\mathbb{Z}+1)\bigcap[2^{\eta-1}, 2^{\eta})$. Select $q_0,...,q_\tau \xleftarrow{\Psi} \mathbb{Z}\bigcap[0, 2^{\gamma}/p)$ with the largest odd integer $q_0$. Select $r_0,...,r_\tau \xleftarrow{\Psi} \mathbb{Z}\bigcap[-2^{\rho}, 2^{\rho}]$, compute $x_0 = q_0 p + 2r_0$ and $x_i = [q_i p + 2r_i]_{x_0}$ for $i \in [\tau]$. Output the public key $pk =< x_0, x_1,..., x_\tau >$ and the secret key $sk =< p >$.

**Encrypt($pk$, $m \in \{0,1\}$).** Select a random subset $T \subseteq [\tau]$ and $r \xleftarrow{\Psi} \mathbb{Z}\bigcap[-2^{\rho}, 2^{\rho}]$, and output ciphertext $c = \left[m + 2r + \sum_{i \in T} x_i\right]_{x_0}$.

**Decrypt($sk, c$).** Output $m' = \left[[c]_p\right]_2$.

To implement fully homomorphic encryption scheme, one applies to it the standard Gentry's bootstappable technique.

### 3.2 Lattice Attack Based on the Public Key

Given a list of approximate multiples of $p$:

$$\{x_i = q_i p + r_i : q_i \in \mathbb{Z}\bigcap[0, 2^{\gamma}/p), r_i \in \mathbb{Z}\bigcap(-2^{\rho}, 2^{\rho})\}_{i=0}^{\tau}, \text{ find } p.$$

Dijk et al. [DGHV10] showed that the security of their FHE scheme is equivalent to solving the approximate GCD problem. Chen and Nguyen [CN11] presented a new AGCD algorithm running in $2^{3\rho/2}$ polynomial-time operations, which is essentially the $3/4$-th root of that of GCD exhaustive search.

According to FHE, we know that an arbitrary ciphertext has general form $c = qp + 2r + m$. The ideal of our attack is very simple, that is, one is how to remove $qp$ in a ciphertext $c$ by adding small noise value. When completing this, it is easy to recover the plaintext bit $m$ in $c$. To do this, we, we define following Diophantine inequality equation problem.

**Definition 3.1. (Diophantine Inequality Equation (DIE)).** Given a list of integers $\{x_i = q_i p + r_i : q_i \in \mathbb{Z}\bigcap[0, 2^{\gamma}/p), r_i \in \mathbb{Z}\bigcap(-2^{\rho}, 2^{\rho})\}_{i=0}^{\tau}$, solve the Diophantine inequality equation $\left|\sum_{i=0}^{\tau} y_i x_i\right| < p/8$ subject to $|y_i| < p/(8\tau 2^{\rho})$ and at least one non-zero $y_i$.

Suppose there is an oracle to solve the above DIE problem, then one can obtain the plaintext bit in an arbitrary ciphertext of FHE [DGHV10]. Since $|y_i| < p/(8\tau 2^{\rho})$, $\left|\sum_{i=0}^{\tau} y_i r_i\right| < p/8$, that is, $\sum_{i=0}^{\tau} y_i x_i$ is only the sum of noise terms, without non-zero multiple of $p$. So, one

can correctly decide the plaintext bit of a ciphertext in FHE according to the parity of $\sum_{i=0}^{\tau} y_i x_i$.

However, it is not difficult to see that the Diophantine inequality equation is a generalization of the knapsack problem. So, there is unlikely an efficient algorithm for general DIE unless P=NP. But, this does not demonstrate that there is not a polynomial time algorithm for special DIE.

To be concrete, we construct a new lattice based on the public key of the FHE [DGHV10]. Given the public key $pk = < x_0, x_1, ..., x_\tau >$ and ciphertext $c$, we randomly choose a subset $T$ from $[\tau]$ such that $|T| = \lambda^3$. Without generality of loss, assume $T = [\lambda^3]$ and $c = qp + 2r + m$ with $|2r| \le 2^\rho$. We construct a new lattice as follows:

$$L = \begin{pmatrix} c & 0 & \cdots & 0 & 0 \\ -x_1 & 1 & \cdots & 0 & 0 \\ \vdots & \vdots & \cdots & \vdots & \vdots \\ -x_{\lambda^3} & 0 & \cdots & 1 & 0 \\ -x_0 & 0 & \cdots & 0 & 1 \end{pmatrix}, L_1 = \begin{pmatrix} c & 1 & 0 & \cdots & 0 & 0 \\ -x_1 & 0 & 1 & \cdots & 0 & 0 \\ \vdots & \vdots & \vdots & \cdots & \vdots & \vdots \\ -x_{\lambda^3} & 0 & 0 & \cdots & 1 & 0 \\ -x_0 & 0 & 0 & \cdots & 0 & 1 \end{pmatrix}.$$

On the one hand, the size of the shortest vector of lattice $L$ is less than $\sqrt{\lambda^3 + 2}\, |c|^{1/(\lambda^3+2)} \approx 2^{\lambda^2}$ according to the parameter setting. On the other hand, there is a non-zero solution $\left|\sum_{i=0}^{\lambda^3} y_i x_i + yc\right| \le 2^{\lambda^2}$ with $|y_i| \le 2^{\lambda^2}$ and $|y| \le 2^{\lambda^2}$ by using pigeon hole principle. This is because $|c|, |x_i| \le 2^{\lambda^5}$, the number of all distinct $y_i, y$ subject to $|y|, |y_i| \le 2^{\lambda^2}$ is $(2^{\lambda^2})^{\lambda^3+2} > 2^{\lambda^5}$, that is, there is at least a non-zero solution for the equation $\left|\sum_{i=0}^{\lambda^3} y_i x_i + yc\right| \le 2^{\lambda^2}$. Thus, if one finds a non-zero small solution vector, then one gets the plaintext bit with probability at least 1/2 ( $y$ is an odd integer).

To conveniently decide, we use a variant lattice $L_1$ of $L$, and call LLL algorithm for lattice $L_1$. Assume $b = (b_0, b_1, ..., b_{\lambda^3+1})$ is the first vector of the $L_1$'s basis output by LLL. If $\|b\|_\infty < p/(8\lambda^3 2^\lambda)$ and $\text{mod}(b_1, 2) = 1$, then $m = \text{mod}(b_0, 2)$. In our experiment, we notice that $b_1$ may be an even integer, but the several vectors following the first vector (such as the second vector, or the third vector, et al.) often satisfy the above condition. That is, the first coordinate of vector is odd and its norm is small. So, as long as one gets one solution of the above form, one can correctly decide plaintext bit. In fact, LLL can also be called many times for distinct subset $T$.

So, we have the following result by applying the block lattice reduction.

**Theroem 3.1.** Suppose the parameters of FHE [DGHV10] $\lambda \le 100$, $\rho = \lambda$, $\eta = 5\lambda^2$, $\gamma = \lambda^5$, and $\tau = \gamma + \lambda$, then there is a running time $2^{\theta\lambda}, (\theta \le 1)$ algorithm recovering plaintext from ciphertext.

**Proof:** According to Theorem 2.1, 2.2, we know $\|b_1\| / \lambda_1(L) \le \sqrt{\gamma_k}(4/3)^{(3k-1)/4} \beta_k^{n/2k-1}$ and $\beta_k \le k^{1.1}$ for all $k \le 100$. If we choose $k = \lambda, n = \lambda^3$, then we get $\|b_1\| \approx \lambda^{1.1 \times \lambda^3 / 2\lambda} \times \lambda_1(L) \approx 2^{3.66\lambda^2} \lambda_1(L) \le 2^{4.66\lambda^2} << 2^\eta$. By using AKS [AKS01, MV10] algorithm, solving each block sub-lattice costs time $2^{\delta\lambda}, \delta < 1$, and the times solving block is at most $\lambda^{O(1)}$. So, the total running time of algorithm is $2^{\theta\lambda}, \theta \le 1$.■

**Theorem 3.2** Suppose the average-case performance of LLL is true, that is, Observation 2.3 holds. Then, for the parameters $\lambda \le 100$, $\rho = \lambda$, $\eta = 4\lambda^2$, $\gamma = \lambda^5$, and $\tau = \gamma + \lambda$, the FHE scheme in [DGHV10] is insecure.

**Proof:** For the above lattice $L_1$, we have

$$\|b\| \le (1.02)^{\lambda^3+2} \lambda(L_1) \le (1.02)^{100\lambda^2+2} \lambda(L_1) \approx 7.2^{\lambda^2} \lambda(L_1) << 2^{4\lambda^2}.■$$

### 3.3 Computational Experiment

In the appendix, we present a toy example to show that our attack method is how to work.

## 4. Improvement

The reason the above lattice attack is successful is that the secret key $p$ is a large integer. If we replace $p$ by a matrix, then the above attack dose not work.

### 4.1 Construction

**Key Generating Algorithm (KeyGen):**

(1) Select a random matrix $T \in Z^{2\times 2}$ with $\|T\|_\infty = 2^{O(\lambda^2)}$ such that $p = \det(T) = 2^{O(\lambda^2)}$ and $p \bmod 2 = 1$. Compute $A \in Z^{2\times 2}$ with $AT = pI$, where $I$ is identity matrix.

(2) Generate $\tau = O(\lambda \log \lambda)$ matrices $\{B_i = (R_i A + 2r_i \bullet I) \bmod p\}_{i=1}^{\tau}$, where $R_i \in \mathbb{Z}_p^{2\times 2}$ is an uniformly random matrix and $|r_i| \le 2^{\lambda}$ and $r_i$ is integer.

(3) Output the public key $pk = (p, B_i, i \in [\tau])$ and the secret key $sk = (p, T)$.

**Encryption Algorithm (Enc).** Given the public key $pk$ and a bit $m \in \{0,1\}$. Evaluate ciphertext $C = (\sum_{i\in[\tau]} k_i B_i + (m+2r)I) \bmod p$ where $|k_i| \le 2^{\lambda}$ and $r$ is integer.

**Add Operation (Add).** Given the public key $pk$ and ciphertexts $C_1, C_2$, output new ciphertext $C = (C_1 + C_2) \bmod p$.

**Multiplication Operation (Mul).** Given the public key $pk$ and ciphertexts $C_1, C_2$, output new ciphertext $C = (C_1 \times C_2) \bmod p$.

**Decryption Algorithm (Dec).** Given the secret key $sk$ and ciphertext $C$, decipher $M = (C \times T) \bmod p \bmod 2$, and the plaintext $m$ is the element $m = M_{1,1}$ of the first row and the first column of $M$.

It is not difficult to verify that the above scheme is a somewhat homomorphic encryption. Now, one can use Gentry's standard bootstrappable technique to implement fully homomorphic encryption.

In addition, we can choose two random primes $p, q = 2^{O(\lambda^2)}$ with $p = a^2 + b^2$ i.e. $p \equiv 1 \bmod 4$. Set $n = pq$ and $T = \begin{pmatrix} a & b \\ -b & a \end{pmatrix}$, $A = \begin{pmatrix} a & -b \\ b & a \end{pmatrix}$ with $AT = \begin{pmatrix} p & 0 \\ 0 & p \end{pmatrix} = pI$.

Now, we can replace $p$ with $n = pq$ in the above scheme, and use the new matrix $A$ to generate the public key $pk = (n, B_i, i \in [\tau])$. We observe that the security of this modification depends on the hardness of factoring $n = pq$.

## 4.2 Efficiency and Security.

**Efficiency:** The size of the public key is $O(\lambda^3 \log \lambda)$, the size of the secret key is $O(\lambda^2)$, the expansion rate of ciphertext to plaintext is $O(\lambda^2)$. To implement FHE, one only needs to add ciphertexts of the secret key to the public key.

**Security:** It is not feasible to use brute force attack by guessing noise term $r$ because $|r| = O(2^{\lambda})$. A possible attack is to solve the following equation

$$\begin{cases} TB_1 = r_1 T \bmod p \\ TB_2 = r_2 T \bmod p \\ \qquad \vdots \\ TB_\tau = r_\tau T \bmod p \end{cases}$$

However this system consists of quadratic equations when $r_i$ is unknown. So, to solve this equation, we also require to guess $r_i$. As well as we know, attacking the above scheme is not feasible by using algebraic equation method.

At the same time, the above scheme can avoid the lattice attack of this paper because the matrix $B_i$ is approximate multiple of the corresponding secret key $A$.

The above improvement scheme has more efficient, but we currently can not reduce its security to solving the secret key.

## 5. Conclusion

This paper presents a heuristic attack for the FHE in [DGHV10] by directly calling LLL algorithm. Our method concentrates on recovering the plaintext in a ciphertext, whereas the attacks considering in [DGHV10] mainly discussed how to avoid to recovering the secret key. Moreover, our attack applies the average-case performance of lattice reduction algorithm, whereas the security of their scheme depends upon the worst-case performance of lattice reduction algorithm.

Our result shows that the FHE scheme in [DGHV10] is not secure for some parameter settings. According to our experiment, one can avoid the above lattice attack by setting parameter $\gamma = \lambda^6$. But, the scheme is less practical in this case.

In addition, we also design an improvement scheme to avoid the above lattice attack.

## Appendix

Here we present a toy example to show the attack processing in this paper.

Assume $\lambda = 3$, $\rho = 3$, $\eta = 3 \times 3^2$, $\gamma = 3^5$, and $\tau = 246$. The secret key is $p = 134217729$. The public key $pk$ is

[1401052789931010491507736140589765585695457925989440123541693166279447146 7
****3250624492796816435996931888772512141226431955536420737938610559094381 4
72****-616680182536192895444061911290757157935072488730318380289539305641 22
66189****389271416239973016638701950598010040551090821895222204461457545694
6680171****729714595409770602191394054738933995667801987076741569193382391 7
88378629****-301404573446266049075794956008230722522321993597035398126086 33
9305890421****-198039757395521807886526608072264557416820515735678022945202
5195354817820****-617125236542825931469309400681208216141848713368988343043
5009695024807072****-314344724456047052407343043408844666369696926387730341
4205945241278676736****-873650667075946688621744141808551515776371375520393
5056180876274225 87406****-566116871703162700785055836208380490355878888679 4
1764453553796826889 10349****303320421668905732839789421362114652076582755 35
18578073761359734805150017****-305996396292740711008757837742538228183237 74
2456329218333166739161 7245338****377518992692665033260320262796931302684111
4587259495227768564459897302000****-602505671843287205560429368109725013582
9510209439555808422352974191399388****676344864456830475824475114988226216 8
4274034250121041986322165411461865 20****432916318934182241896449313215687 26
9091214012863851827110667856719679 1969****366385414764461945789706576688701
4377637049690758932503955703872805828158****644088033922390690810893886486 3
0520345136519463634280039916485130620 02965****-497431137271082186892430790 8
7380507678714523768532323217694400716814289 10****309610755753782621838321 03
5904466679022952996964095569513990204112782637****411359874353223870049675 2
0975910521683746274322735915473823380289155384 47****-296559253157010889630 6
8155424104447246742553396201472546415887068052658 75****30002227770025267391
858910000884871059990143111153411635334237359602604 48****6786897403094134 53
4533013052616306218273301665668499059266421763674152516****-488122982551147
6781567461424022352227982595672304581914582176843266989485****-833738745194

333061366341717042061148501976564139759009870084454456527257****69121095564
1693295523861682704479351414199614493256884144756420136122032 6****372487628
14376402351849927866738412169471758124275191033330375746196819 18****-483098
9279881589641800850063563446860654173481021189969240348597706270293****4343
0100029383653620399194019654636848345442311077613759575762670990848 0****-44
2986670246033315315375636273063608892547901270206390929358242366216 8942****
3085453373300086999038525086562434538237700869467988275252812015174528201**
**129022909858729481661479871168529130260180061799116292527231012665273864*
***-69655158906360626481385109145334455945277321521027925864745111362414757
22****287909033608209074646471105439706662606646843319030070486414901836285
5901****-470279667653424510554457455411828872378527455396028801770918322273
5491445****-509042285041100694838289036102829363655455298886601941040198960
1525964217****4973025390919351667746428837714935897984817309684591162468961
558033108788****131977931889054771511858467543050583792156262260208230411 80
311358711688 48****-29763031934767436025796048947518526262569279851236981623
2857371183389208****261845900414504832028535585117918051658420911464234588 1
759646269617599778****-292799729844852469235858435492459896633589949438016 1
99250526244766579568 6****-472503828108153596940178316032137842152513012586 1
5571190113128080949488 75****-255274146056520933608052986991390274029191918 0
87741918006639555003573211 0****3901699924883018775934201300876403970839443 1
3498114417322911342455342155 5****-446727991025183613831080873861333888597 72
401180162297711052996921308190 04****-84473559971894142356630707378731863974
709865171146196376861479148768593 9****-42956300797501967435794496460485653 3
504736220534428451029666154096166856 0****-3029257715591396344995710588595 16
6721290477519576873874082532569637566811****674326503694919004322821640265 7
609198348578396748570110058646286561550543****4416460957307328933429651971 0
5473441310410352088540233026617207890962900 7****-317401618143181201326262 29
19689487063225445177829741329126617029470565229****-848416514878393781614 13
00423347293523862570865967195448969610406735945 1****-637157417803821906835 7
49309144004499777251837939798014205342319765474252 3****-69811174141255611231
42126304394664100692240845113395656067263535936796 41****-139885310957537355
8878863462248773819221655910678546125803561319483634703****294503765726581 7
814517278448645241121603239592022319035129433362668508455****12408795266068
99322219801421636998955880138912594509904319162569086667857****48689603393 2
55542343800399891318436201582432562616081860784798563390747 04****372115204 8
929805643670945111689388318148008011610863330250146384313490514****10162325
34789170560886537530413112809067305608371593893156782381396974 17****-219125
504396095178106957686965842383565367210007716034357196062680020932 4****-382
814616510531185583457144960985806864456405955342490467358581940678817 6****-
43435027424618024699303342754418427538228299938460064791809869067293758 01**
**-318497000388442771932146182408217788535626694157128314499083373857008998
1****29780831337716239057916317474992943833184110539925800391298910782226 03
4****-35646299698445790397042228305500716607555672552015118265237063836728 0
7036****-117814514750145236247023144463391452133517967475281727880254620941

4597740****-540461964151349087138710622365992939478471801487208382428212328
131439279****44493674969757118022960475250416072148780316595046124594206140
74500866931****-1003192190417074850279007503064750357597362782916327168769 8
8882286796962****5247773905144915577982791469360230589831375237870854888767
524397506943584****-2584569190538017487001826401288424459027320189802441680
5590853343453168 3****1345906862929077728007430595611790983524021827635096 29
964921399828755728 0****6832555530698354154208184847496699552786581776950053
518279338143787925699****4344292501041841247531277937827760067655835298054 6
4653675096351740238283 6****3565549035909548997188530113892754477489762189 90
520607858563844128481917 1****-2072860144697935504054284088099918001014507 94
98797469072524683364624885 50****-6733343462774956196625016076993997323771 99
149247973671636686688741351021 7****-3583299923340827058424254640416287990 94
17097793644789096903689478039542 08****668264399910743855967117937942764960 8
91755104790091634199977176239661086 1****-7377727757231209119527417478030262
652589494862102936983393877670168049 20****3023219858150274908015176264777 58
915170832080798329343123373884475404898 4****24009325106873869239865850455 53
19254350367054768661578071756970885420829 2****383279626589877023218876292 04
874995563003430047659163348782799388693518 22****-42182946718981026744151217
4162689398544803045956933365728781576671046842 7****-347616575081758487291 72
5429694544407427914190058910136682364210900287882 9****-2460987974386609 2092
9211110424073356046208559315193015471725912255958078 9****3183766855011831 32
048185094845057328712298783302213895407080454365003386 2****-29453977389033 0
074818908217086881546433948252341890528335128858831883631 8****589071932867 5
17738521982019356889665255474440059869509833249388287197766 7****-103743821 7
03367089107510450285440788308635114718163849769514562966467841 0****2316123 8
03328243058169233098571879121296068907575335292078207500886192478****-37836
24052802107436036940508974529614791450145443915120710146311154511382****-517
126888035755337636662843257820910955416348532391072713362237024191621 4****7
949479951249130968940991572585647854817499040074945020025733130409999 71****
120825462684048850679659717751994282569739060992240111389659056353371876 5**
**653609994125758412406789620655294705020635717022860027504674628555195774 6
****-615824267591023839830332511270223985058042271795632557126689946022037 5
090****34329110107151794792514133258973658984233479662135949431219423240163
74824****4934642905184971422598474520392179301038186211935247031976375817 50
4175271****-647577701749314883559958894743301728238055335298892032742674286
6741870600****59051354787927436960840187749901290154860626019109939828825 00
157846480771****52496021648614562082212569155496934558359706862960250886779
25341891486724****-2410589420524517936270177224894264402871223938968573947 9
70536614754041444****13002639854530527032622304351899461360474876298518074 0
133207458329715986 8****22238378759998970959512132971272770062311498250530 15
3593439744266033893 1****-107789005967086935450160934159521192229029445998 89
04710660928615743821 8****370267802804320200831629711007978381181523471209
6459791554622717471128055****428311225185672970915242458034974769609057189 0
5102589340751822704189858 38****318477367113117024661666473052340362997653 74

024174792324979201939500136614****-18044366619428829597608810401481706232318
01138791257119437085702886608274****-16353706589928454641656639209430889860
15637374194605760947609411154767191****-34122870423766141314894306174613236
65681462804021888303121320979544895210****-19429030847793062015542223208892
65461714375245771508160044447300727438427****404639194147024293566817148289
876210315875974797747256034596608825277491 9****-16873473958384956282114055 0
6269651756400201409842966154510075174007137888****47948604508570764948833 20
111320196918422144174269953871296363857120826287****275304592708389576138 36
3326166525416389910587673610788696571957764825385 0****-3848063381788017209 8
06858207864190850352215531845558108576432453975165083****505882214698905872
90864782567555125141525890221040364281756291068199703 09****323889577110856 5
951758749062625303658741790055730038514572234190247737603****-1710086495650
4583797626048525967802334811143509535468457133426088667228 65****11799451009
413511871107121252983981050896108278654531729300419918297786 62****346505614
48596586811103471842695819539033430312217993710900252842486779 10****6088148
1005069249692811957480429815280967686204871755252326298910645502 6****-61037
390743738256226566516281557373766646657101213076743768687599386386 81****197
72834347852425860341932257335472182054760983385038442238187417944324 85****2
29294704043218777510163744331696021009622132040312396953657869658205280 1***
*56062543469954728542769717784427630133704118533791752088637714663337223 27*
***-46209687733581641490210473591868387145103083405857003253021586778793034
68****-37014483829595262336700490750122718643702939658862497597053150820103
17451****-10160372489540278260386801643134293111642958947654401961334837688
89209203****45837477529626145053677778083438340105574543689803489949921 5115
4939763088****67313430416438390271924709216914674492127262329818358909372 79
669390502595****-4974792658090714406331898927099583098474530386726606685140
581991524292439****63233245978510138863407444111848213662536091403934283653
38900014943720275****-14066787917294212901682178667499911136957065648450670
184022675250074353 1****-329214852376141981869061601848161883563053216196396
22335357615280742788 80****453781616749338343759274683557183780012746317499 6
89809956278872784999 492****421549114631201287503199373051819100003929952180
629297593165155446318595 8****27030479519718557709929848266071876986472695 90
52748239679892837381435413 3****6177465168960119258539621679913106920411550 0
733089868191141579400634947 10****1442824930255168461970511292973976953917 04
260138098537686781242403806259 5****23949724731746307817499513328519274169 70
3737575978806972825856996913259 15****573185702985916307631845034528175383 32
3814259522824973256217652317111679 41****-1239308670947141849304548530509972 0
07809452814009033964293844428891438 013****-256318951420812055712006577207 96
02481383969017178836132191704706126501 637****-66461277004723628372502928761
43045395841518676566068110238862480478180 958****323879690074601678422397282
39029760364573937233051431448782704345073250 34****-44008503234767508816486 4
5495395177241704183661703089619108749419128555 5****-154736894423063068670 94
169065079273427404971644966549146902961969114961 06****21688531933798212222 5
3757740738405414651881053021300345074360496026961 147****-57842675276894391 5

7616263351822789734387414843562317084463328508977933176****-218129879811312
653610920731196981636006978859540213674648833374422454591****89960316468422
9652540988767016341541079859725348786339094924995367823588****-293073428917
505131737681508307073747950103507564065931617593183449459079****63842592940
09307807592478090601031188097161700896010499984381232052924462****179786967
4327514030310565670256913936930560278916428637693806560628430536****-474623
1159565794461388186646710810043904841533822055413758008688496598133****-290
6195810530300542335582637251423362984618981579006382070478621590151909****5
803998123311226027759698992595849737065395642355716565011370393908954189***
*2693382085931466081508642774365001777542759923744565790931572374148948102*
***-46719558980422136694832905930542718820812079592775128781957070587499863
79****280446693743567128323840872140024300922262436357692336176603624353720
8542****3324002237075738651159067522079172611729239485094504570883372144094
733919****-3911591599851218062191825928015676602570980566358033321130376714
626000703****25965174893214709505842411296486390743544676304955488174623052
19069082221****657191637256559217978148969945375250153490358771709295158007
96197391532****620770789743513142751251834670879173848331838926532630517638
1562658488080****-242562487151159090244959594779093435442850336825725046485
2588504384390314****-411491072229461361520601703991865123616319707934462829
8424776544698953986****6067821307969645430056465136230114987352126895523154
683695781591603561104****-2133677020645229820449576387686129911282190848937
097431272121333119740710****24005916761527698350423525172110297883289733421
31679443493387860326767953****195698877131888537349656570675292201909699391
8539259619253129478137682552****-648560586886890670698888574274215434606119
5149758365163601859518028515971****4404029878818596053651267534637569063233
696226881564778616608783650219983****-1760522782284460375079538981561438012
397545448146653305625381660352583018****-5883697409017386014029072891955594
021866666137100254079977609861752004591****12828209756945805363618176771956
52748244112166039584308488741529018652597****-60068101932628867334730802704
62837805110653229081304629712616836349796659****-96931986428087161387046321
5044898137085063784386093010590483229017918724****-415224682148098051897893
0701022651302347991807869186030271812518549210828****6115049617829066781139
625471598654691279076911284065939077575835082151312****-3582284292535142650
766885041924356001922341663722388170009130693850371655****-6109616276284967
476949882832474617586495754354083557421855161737798447303****34605361857436
6721232145972528107315047339229691188942440901358096489671****-304951663851
2898991055372196992991779682360214374535077965278535423386917****2425082385
720622817683683910704640113429340552003094395090130449181480327****28211792
82639579493514998851690522697278774768379459790006878281644487381****295096
3417697699890060885221159568465896463907591657601823770069647398176****6409
246885112945417236553650396252123773852425925970959900512329872801843****32
45009629173099713491810597084234387399520458204006434715633333171012312****
4002905124328973782650304832296357312635926070389930394295441249877692694**
**-565259019587550919314993354704360075875778475465892081850810356785621919

3****759841291014192185756116379239814123639057616496324042755215291460243856****33253592638915152357373229889666488008471859462403788922842270818389860 72****44546923665324487094391432433306797467438201557786551588439506854152 62350****-660106288484641789187927286878665956020267883362942466613647863266 0447240****37166315949346069241642199995347193397883958747512295307332216 71643782170****123689319223817772962638315968178003594555150162812042159810 2922892453866****161244698271498962232024694832248546560119907562209767046308 9418266878200****-392602381403492191555463048803126315582911466815540514687 0812352980151855****-188694779971260464095893879389244289833590940418011446 7521848195015148824****-682215999784012469382950814873658708047349063401035 41574509145919181 2390****588993023785537556963852725422442074990500599553 920245630488081187 8512547****-529354683958828669084944332624003260613184147 3650402241738505495673077236****-597917361105885896578491554967051396265145 2427945412139134271212133925614****-159343645732901543526315534035601053662 2218309604765106178616517680381 19****1000840131863824075770467970823941585 1 85311537326121174050912065708832666****526478771285427649824712845833777249 1398965534209903682821389603795028343****-672005513047781634369956318992112 919863392381111098506908408387508847473 1****294856402660626487909290039263 4 8806199078021689219595106844103853913491 91****389145792553946364192545458 47 58481567622192776097758268275613780169729305****334447473216037956868904506 99957996555090672376924459248399413647581451 42****-571376066176687918631997 2299683530585628527838025651917155479895225872620****-581348793934927237844 1256727908004257002812844400683832807978680450901433****-324366020689897318 737581529753651025461030321861417386928076212505437158 5****-644169259364777 072925053382696809813312596672341900841775348306486264328 2****-435150649382 13138937425329095829862460273878220702478624289006581505497 77****-355915415 871103319316100022994810127794771386069767516638424761664717032 2****-557896 979433143713802833291565330996553308916612291742740961673268577194 8****-473 4714094685487338147886763963121855252450865430203756011018865861908853****- 2428201399383130458681275708372397922112027158877311222049557118076158914** **507943545594478663358665582594379210824016330119623677575213765931021481 3 ****-387445736615939025677968262437568373137293009326147239083334851256365 9 174****-157624825918230608097543547256863989930416322655898656322527622800 9 318114****-361785994941507422016627539975913584679770462582948289744826116 1 603678974****31886013139486422625440274143822924080931402871626741214996597 60607357032****5730625864116901675836130067060734822408929079134589881541 48 0046771543753****-553659386145736494235324490205917552612265056124108931785 7772311113652636****153382585639839074569609697882370743907352885865011274 6 7123004833701592****446689057280918031323658943720130674408154034539191060 8 877098548181826460****467825613189753147390140074844198004300863929369093 97 0851218324996860708 7****-525037559493108248724202792845450221836415905751 08 2652142529765936980150 7****434196090289962138013330748854830541146285664207 325579855076365772169534 9****-502793245900738819100099657399322888386874331 0699584444900112214668799970****7382106921817153500831697383062242938121595

91627831190636770563222467121****-38572970228695463135441526795671827926051
49427110103780679557956298703351****1100683504620682509436890715542655705 61
26904165505209443387475267509071 64****256701261435954569910392535564578835 8
0339769361135121633464748303341 95259****-18017666427427347764701130010108 44
47069812004225316541640778778606 0297744****16460595207280505429710646891864
21708561004359998173781458855498948087914****-180694112449940488347478195 99
842692851409554134840350685029968791 6597745]

The lattice $L_1$ is as follows.

C=
[[-1968487892819738597274658441513155537250551194506972917051476635672423 73 1
0 0 0 0 0 0 0 0 0 0 0 0 0 0 0 0 0 0 0 0 0 0 0 0 0 0 0 0]
[-32506244927968164359969318887725121412264319555364207379386105590943814 72
0 1 0 0 0 0 0 0 0 0 0 0 0 0 0 0 0 0 0 0 0 0 0 0 0 0 0 0 0]
[6166801825361928954440619112907571579350724887303183802895393056412266189 0
0 1 0 0 0 0 0 0 0 0 0 0 0 0 0 0 0 0 0 0 0 0 0 0 0 0 0 0]
[-38927141623997301663870195059801004055109082189522220446145754569466801 71
0 0 0 1 0 0 0 0 0 0 0 0 0 0 0 0 0 0 0 0 0 0 0 0 0 0 0 0 0]
[-7297145954097706021913940547389339956678019870767415691933823917883786 29 0
0 0 0 1 0 0 0 0 0 0 0 0 0 0 0 0 0 0 0 0 0 0 0 0 0 0 0 0]
[30140457344626604907579495600823072252232199359703539812608633930589042 1 0
0 0 0 0 1 0 0 0 0 0 0 0 0 0 0 0 0 0 0 0 0 0 0 0 0 0 0 0]
[1980397573955218078865266080722645574168205157356780229452025195354817820 0
0 0 0 0 0 1 0 0 0 0 0 0 0 0 0 0 0 0 0 0 0 0 0 0 0 0 0 0]
[6171252365428259314693094006812082161418487133689883430435009695024807072 0
0 0 0 0 0 0 1 0 0 0 0 0 0 0 0 0 0 0 0 0 0 0 0 0 0 0 0 0]
[3143447244560470524073430434088446663696969263877303414205945241278676736 0
0 0 0 0 0 0 0 1 0 0 0 0 0 0 0 0 0 0 0 0 0 0 0 0 0 0 0 0]
[873650667075946688621744141808551515776371375520393505618087627422587406 0
0 0 0 0 0 0 0 0 1 0 0 0 0 0 0 0 0 0 0 0 0 0 0 0 0 0 0 0]
[5661168717031627007850558362083804903558788886794176445355379682688910349 0
0 0 0 0 0 0 0 0 0 1 0 0 0 0 0 0 0 0 0 0 0 0 0 0 0 0 0 0]
[-3033204216689057328397894213621146520765827553518578073761359734805150017
0 0 0 0 0 0 0 0 0 0 0 1 0 0 0 0 0 0 0 0 0 0 0 0 0 0 0 0 0]
[3059963962927407110087578377425382281832377424563292183331667391617245338 0
0 0 0 0 0 0 0 0 0 0 0 1 0 0 0 0 0 0 0 0 0 0 0 0 0 0 0 0]
[-3775189926926650332603202627969313026841114587259495227768564459897302000
0 0 0 0 0 0 0 0 0 0 0 0 0 1 0 0 0 0 0 0 0 0 0 0 0 0 0 0 0]
[6025056718432872055604293681097250135829510209439555808422352974191399388 0
0 0 0 0 0 0 0 0 0 0 0 0 0 1 0 0 0 0 0 0 0 0 0 0 0 0 0 0]
[-6763448644568304758244751149882262168427403425012104198632216541146186520
0 0 0 0 0 0 0 0 0 0 0 0 0 0 0 1 0 0 0 0 0 0 0 0 0 0 0 0 0]
[-4329163189341822418964493132156872690912140128638518271106678567196791969
0 0 0 0 0 0 0 0 0 0 0 0 0 0 0 0 1 0 0 0 0 0 0 0 0 0 0 0 0]

[-3663854147644619457897065766887014377637049690758932503955703872805828158
0 0 0 0 0 0 0 0 0 0 0 0 0 0 0 0 0 1 0 0 0 0 0 0 0 0 0 0]
[-6440880339223906908108938864863052034513651946363428003991648513062002965
0 0 0 0 0 0 0 0 0 0 0 0 0 0 0 0 0 0 1 0 0 0 0 0 0 0 0 0]
[4974311372710821868924307908738050767871452376853232321769440071681428910 0
0 0 0 0 0 0 0 0 0 0 0 0 0 0 0 0 0 0 1 0 0 0 0 0 0 0 0]
[-3096107557537826218383210359044666790229529969640955695139902041127826370 0
0 0 0 0 0 0 0 0 0 0 0 0 0 0 0 0 0 0 0 1 0 0 0 0 0 0 0]
[-4113598743532238700496752097591052168374627432273591547382338028915538447
0 0 0 0 0 0 0 0 0 0 0 0 0 0 0 0 0 0 0 0 0 1 0 0 0 0 0 0]
[2965592531570108896306815542410444724674255339620147254641588706805265875 0
0 0 0 0 0 0 0 0 0 0 0 0 0 0 0 0 0 0 0 0 0 1 0 0 0 0 0]
[-3000222777002526739185891000088487105999014311115341163533423735960260448
0 0 0 0 0 0 0 0 0 0 0 0 0 0 0 0 0 0 0 0 0 0 0 1 0 0 0 0]
[-6786897403094134534533013052616306218273301665668499059266421763674152516
0 0 0 0 0 0 0 0 0 0 0 0 0 0 0 0 0 0 0 0 0 0 0 0 1 0 0 0]
[4881229825511476781567461424022352227982595672304581914582176843266989485 0
0 0 0 0 0 0 0 0 0 0 0 0 0 0 0 0 0 0 0 0 0 0 0 0 1 0 0]
[8337387451943330613663417170420611485019765641397590098700844544565272570 0
0 0 0 0 0 0 0 0 0 0 0 0 0 0 0 0 0 0 0 0 0 0 0 0 0 1 0]
[-14010527899310104915077361405897655856954579259894401235416931662794471467
0 0 0 0 0 0 0 0 0 0 0 0 0 0 0 0 0 0 0 0 0 0 0 0 0 0 0 1]
]

By calling LLL algorithm, the reduced basis of $L_1$ is

B=[[-86 122 -65 -175 -90 -182 113 79 41 46 -225 99 -72 164 -66 -376 5 -55 167 -159 94 96 33 -63 -1 -42 -39 -92 0]
[-87 -49 65 -321 -209 49 11 -30 29 48 -149 181 12 109 -153 -237 -43 -83 10 79 177 -120 -127 171 17 100 -89 -52 4]
[175 75 -43 -80 36 -86 14 -147 -111 -180 -60 -5 -181 308 -98 -114 115 -96 150 -151 184 293 48 -39 2 8 57 52 -4]
[-153 -21 -61 172 138 198 -31 -188 -3 107 61 47 260 42 30 -55 -82 64 -91 -52 -31 179 -59 -104 -113 72 -25 -6 9]
[77 149 -12 60 242 89 212 23 90 126 73 -40 56 -135 91 -49 -68 -8 116 103 100 91 100 80 -55 -114 57 -45 -5]
[-169 55 -115 362 140 102 -157 23 -69 84 -9 4 145 4 5 97 110 -113 -22 76 -59 -83 34 -88 -71 107 9 39 14]
[-23 -143 -137 54 -184 7 -209 32 -67 234 -9 179 345 6 -7 -109 -143 40 -2 89 -164 -110 -109 -11 -80 128 -48 79 18]
[-62 -66 -64 -232 64 131 1 -175 -42 -107 -145 170 26 234 -154 -95 119 124 -128 -281 211 111 55 -82 -7 91 -68 -87 -38]
[-155 167 -110 86 -19 -102 96 108 120 178 -113 33 -161 -32 -9 -187 -33 -62 145 66 87 -149 -39 -96 176 62 -115 -206 10]
[-31 -5 56 2 97 146 -42 -213 -88 -2 -173 -99 74 214 -64 -53 -50 -156 -16 -51 21 96 -244 150

-60 -31 -53 157 85]
[-56 40 -21 109 -73 -140 -97 3 -28 -255 2 -59 -10 -161 196 2 -14 -76 242 -66 -33 60 3 -19 -136 -66 119 69 -14]
[-223 -35 50 -147 5 -171 -72 52 3 94 -53 103 -4 204 -69 -250 -76 66 -56 79 -28 23 -256 -68 24 21 69 10 9]
[144 -22 -83 -257 -39 -19 16 -39 -131 64 -34 -75 -137 11 -97 76 9 -168 -214 89 64 -125 -8 -189 52 34 28 20 -38]
[-168 98 -91 42 18 -101 365 217 -31 -108 -110 62 14 -63 70 -9 45 -70 -129 91 108 -34 89 38 -85 10 -110 -162 -4]
[-146 -28 -46 -3 7 -61 197 106 -149 -57 -17 -77 57 2 -74 147 19 -23 -98 223 -120 -166 58 -69 -130 -63 177 -90 -44]
[84 42 137 -208 195 108 130 8 72 16 -40 -25 9 -102 -114 43 -115 78 7 97 39 -272 -52 -87 -181 -136 60 -19 -6]
[-169 3 -26 -42 50 -16 4 222 184 224 -115 202 -127 -97 21 -88 198 53 121 88 11 -81 83 60 105 38 48 -55 -43]
[-15 -105 183 181 -118 53 -54 39 51 56 -63 -106 -43 14 56 153 -43 103 140 -99 -207 -63 -129 -100 32 -45 -122 -72 35]
[186 16 12 -98 126 -94 45 37 -140 -12 -16 68 -26 240 -18 30 -121 47 168 127 21 -25 -51 154 -151 -16 -23 -35 -5]
[-31 -5 -119 190 -1 -34 -9 126 -23 34 103 104 86 -82 55 -60 -127 106 29 43 -53 -1 -118 11 115 136 38 86 47]
[224 110 -166 50 225 142 -73 -94 29 38 77 -84 9 51 -127 83 -74 16 154 9 -5 53 237 15 65 -8 154 -52 3]
[-16 12 93 -44 16 319 -146 -30 -26 88 118 124 112 41 -47 -134 6 -130 -56 96 136 90 77 174 -19 69 48 -128 -16]
[25 -25 -142 43 -65 -23 54 -45 -159 -148 118 103 143 46 145 -223 -107 27 72 21 88 148 -72 21 -54 62 40 17 -79]
[-35 -45 -4 27 -343 -109 -73 32 62 -25 -196 76 118 -39 26 -241 -147 132 198 -112 -90 -10 122 -113 -126 -137 -51 -31 25]
[47 -69 4 85 -139 -116 90 148 81 -221 -62 -172 86 -206 126 323 8 266 -45 -106 -136 -123 163 100 -120 -51 15 -132 9]
[129 -13 -17 100 360 214 -2 -63 -90 23 -68 -87 53 -157 14 181 31 100 28 87 130 -87 -111 -22 46 7 146 -32 -99]
[-201 -65 -109 -13 -128 -179 -83 50 -60 56 109 105 -12 51 35 -111 -18 242 19 -119 -109 230 2 3 1 -33 -85 -11 -12]
[32662 1532013 35166 -334620 -492845 319870 -62472 -112310 -73327 -101190 -187515 444100 363631 224003 356632 512681 263715 351591 -34152 266919 -280216 127712 -299356 -168344 363922 -258533 45283 138299 -195047] ]
When calling LLL algorithm, generating matrix *U* is as follows.
U= [ [122 -65 -175 -90 -182 113 79 41 46 -225 99 -72 164 -66 -376 5 -55 167 -159 94 96 33 -63 -1 -42 -39 -92 0]
[-49 65 -321 -209 49 11 -30 29 48 -149 181 12 109 -153 -237 -43 -83 10 79 177 -120 -127 171 17 100 -89 -52 4]
[75 -43 -80 36 -86 14 -147 -111 -180 -60 -5 -181 308 -98 -114 115 -96 150 -151 184 293 48

-39 2 8 57 52 -4]
[-21 -61 172 138 198 -31 -188 -3 107 61 47 260 42 30 -55 -82 64 -91 -52 -31 179 -59 -104 -113 72 -25 -6 9]
[149 -12 60 242 89 212 23 90 126 73 -40 56 -135 91 -49 -68 -8 116 103 100 91 100 80 -55 -114 57 -45 -5]
[55 -115 362 140 102 -157 23 -69 84 -9 4 145 4 5 97 110 -113 -22 76 -59 -83 34 -88 -71 107 9 39 14]
[-143 -137 54 -184 7 -209 32 -67 234 -9 179 345 6 -7 -109 -143 40 -2 89 -164 -110 -109 -11 -80 128 -48 79 18]
[-66 -64 -232 64 131 1 -175 -42 -107 -145 170 26 234 -154 -95 119 124 -128 -281 211 111 55 -82 -7 91 -68 -87 -38]
[167 -110 86 -19 -102 96 108 120 178 -113 33 -161 -32 -9 -187 -33 -62 145 66 87 -149 -39 -96 176 62 -115 -206 10]
[-5 56 2 97 146 -42 -213 -88 -2 -173 -99 74 214 -64 -53 -50 -156 -16 -51 21 96 -244 150 -60 -31 -53 157 85]
[40 -21 109 -73 -140 -97 3 -28 -255 2 -59 -10 -161 196 2 -14 -76 242 -66 -33 60 3 -19 -136 -66 119 69 -14]
[-35 50 -147 5 -171 -72 52 3 94 -53 103 -4 204 -69 -250 -76 66 -56 79 -28 23 -256 -68 24 21 69 10 9]
[-22 -83 -257 -39 -19 16 -39 -131 64 -34 -75 -137 11 -97 76 9 -168 -214 89 64 -125 -8 -189 52 34 28 20 -38]
[98 -91 42 18 -101 365 217 -31 -108 -110 62 14 -63 70 -9 45 -70 -129 91 108 -34 89 38 -85 10 -110 -162 -4]
[-28 -46 -3 7 -61 197 106 -149 -57 -17 -77 57 2 -74 147 19 -23 -98 223 -120 -166 58 -69 -130 -63 177 -90 -44]
[42 137 -208 195 108 130 8 72 16 -40 -25 9 -102 -114 43 -115 78 7 97 39 -272 -52 -87 -181 -136 60 -19 -6]
[3 -26 -42 50 -16 4 222 184 224 -115 202 -127 -97 21 -88 198 53 121 88 11 -81 83 60 105 38 48 -55 -43]
[-105 183 181 -118 53 -54 39 51 56 -63 -106 -43 14 56 153 -43 103 140 -99 -207 -63 -129 -100 32 -45 -122 -72 35]
[16 12 -98 126 -94 45 37 -140 -12 -16 68 -26 240 -18 30 -121 47 168 127 21 -25 -51 154 -151 -16 -23 -35 -5]
[-5 -119 190 -1 -34 -9 126 -23 34 103 104 86 -82 55 -60 -127 106 29 43 -53 -1 -118 11 115 136 38 86 47]
[110 -166 50 225 142 -73 -94 29 38 77 -84 9 51 -127 83 -74 16 154 9 -5 53 237 15 65 -8 154 -52 3]
[12 93 -44 16 319 -146 -30 -26 88 118 124 112 41 -47 -134 6 -130 -56 96 136 90 77 174 -19 69 48 -128 -16]
[-25 -142 43 -65 -23 54 -45 -159 -148 118 103 143 46 145 -223 -107 27 72 21 88 148 -72 21 -54 62 40 17 -79]
[-45 -4 27 -343 -109 -73 32 62 -25 -196 76 118 -39 26 -241 -147 132 198 -112 -90 -10 122 -113 -126 -137 -51 -31 25]
[-69 4 85 -139 -116 90 148 81 -221 -62 -172 86 -206 126 323 8 266 -45 -106 -136 -123 163

100 -120 -51 15 -132 9]
[-13 -17 100 360 214 -2 -63 -90 23 -68 -87 53 -157 14 181 31 100 28 87 130 -87 -111 -22 46 7 146 -32 -99]
[-65 -109 -13 -128 -179 -83 50 -60 56 109 105 -12 51 35 -111 -18 242 19 -119 -109 230 2 3 1 -33 -85 -11 -12]
[1532013 35166 -334620 -492845 319870 -62472 -112310 -73327 -101190 -187515 444100 363631 224003 356632 512681 263715 351591 -34152 266919 -280216 127712 -299356 -168344 363922 -258533 45283 138299 -195047] ]
The above three matrices is satisfied to equality U*C=B. Moreover, U is equal to B except for the first column.
Now, we can decide the plaintext bit in the ciphertext
-19684878928197385972746584415131555372505511945069729170514766356724 2373
according to the parity of the first column of U and B.
It is easy to check that they are respectively

[0 1 1 1 1 1 1 0 1 1 0 1 0 0 0 0 1 1 0 1 0 0 1 1 1 1 1 1],

[0 1 1 1 1 1 1 0 1 1 0 1 0 0 0 0 1 1 0 1 0 0 1 1 1 1 1 0].

So, the plaintext is "1" for the above ciphertext. This is because the first columns in U and B have same parity if the plaintext is "1" in a ciphertext and $\|U\|_\infty, \|B\|_\infty < 2^{\lambda^2}$.

Notice that the last row vector in U is too large ( that is $|y|, |y_i| > 2^{\lambda^2}$ ), so the last terms in the parity vectors is not satisfied the above condition.
On the other hand, suppose the ciphertext is
-19684878928197385972746584415131555372505511945069729170514766356724 2374,
then calling LLL generates the matrices B, U as follows.
B=[[-110 112 -87 -84 7 1 161 -66 239 63 -181 -146 -205 80 -74 -63 37 -41 -18 34 85 75 7 106 122 158 -27 45 1]
[-2 5 73 41 131 -131 -125 153 -181 217 62 166 201 -63 -140 15 42 36 60 8 -148 -1 96 122 -24 -46 149 170 1]
[102 -33 -62 9 73 -207 127 -42 -273 170 1 130 185 164 -30 -172 -66 22 20 -128 -109 -132 -110 -184 59 -71 84 122 1]
[68 -152 186 187 -215 -37 129 59 14 -153 180 40 -52 203 6 88 139 96 -195 70 -129 -308 -57 -56 139 78 -65 -48 0]
[48 62 -104 -173 250 -14 -52 -73 -173 -23 173 -12 145 44 -217 -93 -62 152 -74 44 210 26 -25 -155 -149 -166 172 167 1]
[88 -57 169 30 -189 7 168 125 26 188 -254 -7 -79 60 -104 -38 133 121 -103 -52 -127 29 -138 318 52 188 -111 58 -1]
[-84 76 222 155 -108 -26 -197 25 -224 297 19 -53 77 -58 5 66 -51 -106 88 -73 -166 -13 37 22 -175 26 -41 -158 -4]
[76 124 -50 4 26 50 49 11 -199 159 -151 -101 -27 6 -104 -149 -14 -201 66 -222 -130 73 -150 -68 33 -27 4 -273 0]
[-94 -37 -17 -71 -51 -45 -65 -68 -89 -85 68 209 52 -21 85 -166 -81 111 -100 -162 43 -4 -175 83 -53 150 -106 143 -1]
[52 -22 64 -80 -114 -107 -63 231 71 -89 -26 108 -215 163 -112 -141 7 10 -78 36 -188 -41 -64

-1 85 95 -40 88 1]
[-34 179 157 24 -63 12 -162 80 -57 -121 56 41 -36 -255 52 -139 -70 -116 24 -92 -60 21 138 130 -209 -65 -12 -225 -28]
[64 100 57 -25 0 35 -36 -82 -131 -40 115 -220 -85 179 -128 -129 -111 -56 -74 -61 48 -146 -60 -55 181 -63 -31 -13 -27]
[-4 2 80 -34 162 74 -169 179 -119 103 -21 -57 28 110 -103 103 -52 141 61 57 165 74 150 -70 -48 -130 89 196 29]
[62 -80 113 181 -168 -152 141 52 -279 -49 32 95 104 60 135 9 99 39 -107 -73 73 -170 -131 3 34 67 -26 -148 -3]
[132 -26 -46 43 -100 102 -85 134 130 -106 49 -5 -41 21 -251 30 130 104 -137 28 -94 -57 -150 78 -12 123 -94 -47 62]
[-14 27 240 137 29 -36 -56 147 -70 94 22 -133 17 -29 -210 193 139 267 -19 42 122 -1 -72 160 -44 39 62 40 34]
[36 8 80 103 -99 -167 -275 13 -142 210 85 50 82 68 106 32 73 -91 54 -91 -63 -195 85 87 63 143 28 48 -23]
[-54 -29 -138 31 -13 35 -94 -49 276 -20 -35 -77 -72 -74 293 46 7 -12 73 112 144 116 -53 91 64 36 1 -89 40]
[50 -15 -12 -100 42 -124 83 -70 98 19 91 31 -120 174 17 -96 109 175 -178 22 30 7 -108 89 -70 -4 -7 207 47]
[-16 -65 106 97 -79 -133 -87 42 43 161 179 185 48 -58 66 17 128 -71 -40 21 -273 -30 196 -89 38 27 60 27 -51]
[-16 20 190 14 2 -83 -192 111 -232 65 9 210 -12 72 -22 -60 3 79 7 -95 -131 136 -20 237 -182 41 25 -180 13]
[-8 -70 -124 17 -73 262 -62 138 3 3 -158 -42 72 -120 203 -14 221 -154 121 -97 148 314 -103 46 -83 53 0 -172 -44]
[-90 -41 -50 -64 -141 85 -20 164 190 -6 -1 5 -156 -46 7 90 34 79 -34 139 60 -60 -35 234 22 46 -119 42 107]
[110 1 107 54 -158 104 -96 -198 63 43 -81 -218 -101 -208 286 32 -121 35 36 -53 -81 163 91 77 -209 -178 -5 -80 9]
[-202 114 -93 1 164 -87 236 -150 147 -19 -82 42 21 156 6 -193 33 24 -38 -147 94 -91 3 -38 53 -76 -11 -13 82]
[-118 212 103 23 -78 -23 -224 -36 124 -62 94 -27 -185 73 -147 -125 -68 -12 -41 116 188 37 216 71 -53 163 85 64 51]
[-38 -34 288 71 -145 -145 -124 170 -21 74 202 50 76 -31 97 62 -80 64 -34 10 -89 14 -70 214 -34 38 -80 90 75]
[31795 1529484 34164 -326891 -481784 312328 -61101 -109858 -71653 -98959 -183586 434089 355349 219063 348775 501243 257557 343487 -33138 260672 -273839 124941 -292486 -164591 355578 -252847 44112 135082 -190712] ]

U=[[112 -87 -84 7 1 161 -66 239 63 -181 -146 -205 80 -74 -63 37 -41 -18 34 85 75 7 106 122 158 -27 45 1]
[5 73 41 131 -131 -125 153 -181 217 62 166 201 -63 -140 15 42 36 60 8 -148 -1 96 122 -24 -46 149 170 1]
[-33 -62 9 73 -207 127 -42 -273 170 1 130 185 164 -30 -172 -66 22 20 -128 -109 -132 -110

-184 59 -71 84 122 1]
[-152 186 187 -215 -37 129 59 14 -153 180 40 -52 203 6 88 139 96 -195 70 -129 -308 -57 -56 139 78 -65 -48 0]
[62 -104 -173 250 -14 -52 -73 -173 -23 173 -12 145 44 -217 -93 -62 152 -74 44 210 26 -25 -155 -149 -166 172 167 1]
[-57 169 30 -189 7 168 125 26 188 -254 -7 -79 60 -104 -38 133 121 -103 -52 -127 29 -138 318 52 188 -111 58 -1]
[76 222 155 -108 -26 -197 25 -224 297 19 -53 77 -58 5 66 -51 -106 88 -73 -166 -13 37 22 -175 26 -41 -158 -4]
[124 -50 4 26 50 49 11 -199 159 -151 -101 -27 6 -104 -149 -14 -201 66 -222 -130 73 -150 -68 33 -27 4 -273 0]
[-37 -17 -71 -51 -45 -65 -68 -89 -85 68 209 52 -21 85 -166 -81 111 -100 -162 43 -4 -175 83 -53 150 -106 143 -1]
[-22 64 -80 -114 -107 -63 231 71 -89 -26 108 -215 163 -112 -141 7 10 -78 36 -188 -41 -64 -1 85 95 -40 88 1]
[179 157 24 -63 12 -162 80 -57 -121 56 41 -36 -255 52 -139 -70 -116 24 -92 -60 21 138 130 -209 -65 -12 -225 -28]
[100 57 -25 0 35 -36 -82 -131 -40 115 -220 -85 179 -128 -129 -111 -56 -74 -61 48 -146 -60 -55 181 -63 -31 -13 -27]
[2 80 -34 162 74 -169 179 -119 103 -21 -57 28 110 -103 103 -52 141 61 57 165 74 150 -70 -48 -130 89 196 29]
[-80 113 181 -168 -152 141 52 -279 -49 32 95 104 60 135 9 99 39 -107 -73 73 -170 -131 3 34 67 -26 -148 -3]
[-26 -46 43 -100 102 -85 134 130 -106 49 -5 -41 21 -251 30 130 104 -137 28 -94 -57 -150 78 -12 123 -94 -47 62]
[27 240 137 29 -36 -56 147 -70 94 22 -133 17 -29 -210 193 139 267 -19 42 122 -1 -72 160 -44 39 62 40 34]
[8 80 103 -99 -167 -275 13 -142 210 85 50 82 68 106 32 73 -91 54 -91 -63 -195 85 87 63 143 28 48 -23]
[-29 -138 31 -13 35 -94 -49 276 -20 -35 -77 -72 -74 293 46 7 -12 73 112 144 116 -53 91 64 36 1 -89 40]
[-15 -12 -100 42 -124 83 -70 98 19 91 31 -120 174 17 -96 109 175 -178 22 30 7 -108 89 -70 -4 -7 207 47]
[-65 106 97 -79 -133 -87 42 43 161 179 185 48 -58 66 17 128 -71 -40 21 -273 -30 196 -89 38 27 60 27 -51]
[20 190 14 2 -83 -192 111 -232 65 9 210 -12 72 -22 -60 3 79 7 -95 -131 136 -20 237 -182 41 25 -180 13]
[-70 -124 17 -73 262 -62 138 3 3 -158 -42 72 -120 203 -14 221 -154 121 -97 148 314 -103 46 -83 53 0 -172 -44]
[-41 -50 -64 -141 85 -20 164 190 -6 -1 5 -156 -46 7 90 34 79 -34 139 60 -60 -35 234 22 46 -119 42 107]
[1 107 54 -158 104 -96 -198 63 43 -81 -218 -101 -208 286 32 -121 35 36 -53 -81 163 91 77 -209 -178 -5 -80 9]
[114 -93 1 164 -87 236 -150 147 -19 -82 42 21 156 6 -193 33 24 -38 -147 94 -91 3 -38 53 -76

-11 -13 82]
[212 103 23 -78 -23 -224 -36 124 -62 94 -27 -185 73 -147 -125 -68 -12 -41 116 188 37 216 71 -53 163 85 64 51]
[-34 288 71 -145 -145 -124 170 -21 74 202 50 76 -31 97 62 -80 64 -34 10 -89 14 -70 214 -34 38 -80 90 75]
[1529484 34164 -326891 -481784 312328 -61101 -109858 -71653 -98959 -183586 434089 355349 219063 348775 501243 257557 343487 -33138 260672 -273839 124941 -292486 -164591 355578 -252847 44112 135082 -190712] ]
Similarly, the above three matrices is satisfied to equality U*C=B.
It is easy to check that the parity of the first columns of B and U are respectively

[0 0 0 0 0 0 0 0 0 0 0 0 0 0 0 0 0 0 0 0 0 0 0 0 0 0 0 1]

[0 1 1 0 0 1 0 0 1 0 1 0 0 0 0 1 0 1 1 1 0 0 1 1 0 0 0 0]

Thus, the plaintext bit is “0” in the ciphertext. Because the parity of the first column of B is “0” except its last row and is different from the parity of the first column of U.

Similarly, the last row vector in U is too large ( that is $|y|,|y_i| > 2^{\lambda^2}$ ), so the last terms in the parity vectors is not satisfied the above condition.